\documentstyle[epsf,aps]{revtex}
\begin{document}
\title{A practically feasible 
entanglement assisted quantum key distribution protocol
}
\author{Xiang-Bin Wang\thanks{email: wang$@$qci.jst.go.jp} 
\\
        Imai Quantum Computation and Information project, ERATO, Japan Sci. and Tech. Corp.\\
Daini Hongo White Bldg. 201, 5-28-3, Hongo, Bunkyo, Tokyo 113-0033, Japan}

\maketitle 
\begin{abstract}
We give an entanglement assisted scheme for quantum key distribution.
The scheme requires the maximally entangled 2-qubit state but does not require
any quantum storage. The protocol is unconditionally secure under whatever
type of Eve's attack.
The threshold of the channel error rate for the protocol to produce larger 
than zero final bit depends on the noisy pattern. In particular, 
given the symmetric channel with independent errors,
our scheme  can tolerate a bit error rate up to
26\% in the 4-state case and  30\% in the 
6-state case, respectively. These values are higher than those of
all currently known two-level-state schemes without using a quantum storage.
\end{abstract}
\section{Introduction}
Due to the Heisenberg uncertainty principle,
quantum key distribution (QKD) is different from classical cryptography in 
that 
an unknown quantum state is  in principle not known  unless it is 
disturbed, rather 
than the conjectured difficulty of computing certain functions.
The first published protocol, proposed in 1984~\cite{BB}, 
is called BB84 after its inventors (C. H. Bennett and G. Bras\-sard.) For a 
history of the subject, one may see e.g. \cite{gisin}.
In this protocol, the participants (Alice and Bob)
wish to agree on a secret key about which no
eavesdropper (Eve) can obtain significant information.  
Alice sends each bit of
the secret key in one of a set of conjugate bases which 
Eve does not know, and this key is protected by the impossibility of
measuring the state of a quantum system simultaneously in 
two conjugate bases. Since then, studies on QKD are extensive.

Most of the entanglement assisted QKD protocols such as the Lo-Chau protocol\cite{qkd}
are not so practical compared
with the prepare-and-measure protocols such as BB84\cite{BB}, 
6-state protocol\cite{6state}, the Gottesman-Lo protocol\cite{gl,chau} and so on because
normally the entanglement assisted protocols require a quantum storage which is
generally believed to be technically difficult. However, producing the maximally
entangled pairs is not a problem by our current technology. Maximally
entangled pairs in polarization space can be robustly produced by the type two spontaneous 
parametric down conversion(SPDC)\cite{para}. The proposed QKD protocol in this work does not require
any quantum storage, though it uses the entanglement pairs and the collective measurement.
The collective measurement can be done by using a polarizing beam splitter as we shall show 
in the section of ``experimental realization''.  
Therefore it is practically implementable by the currently existing technology.

Normally, the channel for quantum bit transmission is noisy. The error to a qubit caused
by the channel noise can be divided into $\sigma_x=\left(\begin{array}{cc}0&1\\1&0\end{array}\right), 
\sigma_z\left(\begin{array}{cc}1&0\\0&-1\end{array}\right), 
\sigma_y=\left(\begin{array}{cc}0&-i\\i&0\end{array}\right)$ errors,
which represent for a bit flip error only, a phase flip error only and both error, respectively.
The detected bit (or phase) flip error rate is the summation of $\sigma_x$  (or $\sigma_z$)
error rate and $\sigma_y$ error rate.
The most natural noisy channel  is the symmetric channel, i.e., 
the   $\sigma_x, \sigma_z, \sigma_y$ errors
are equally distributed. One can take advantage of this type of symmetric channel in the QKD protocol\cite{gl}.
To be sure that the channel noise is symmetric, one must use 6-state protocol. In the BB84 protocol, since the $\sigma_y$
error can never be detected or deduced, we have to assume the worst case of zero $\sigma_y$ error\cite{gl}. This is why
the 4-state protocol can only tolerate a lower error rate than 6-state protocol. This is the reason that
the largest tolerable bit flip error rate or phase flip error rate for BB84 protocol is $25\%$, lower than 
that of 6-state protocol, which is $33.3\%$\cite{gl}. However, in our entanglement assisted protocol,
the channel flipping rate upper bound of $25\%$ for the 4-state protocol is broken. 
The tolerable channel bit-flip and phase-flip rate is raised to $26\%$ for the symmetric
channel. This is not a strange result if we consider an entanglement purification protocol 
with bit flip error rejection
and phase flip error rejection alternately: even though $\sigma_y$ error is never detected or deduced, the tolerable
bit flip rate or phase flip rate can be as high as $33.3\%$ for the entanglement distillation. In such a case one
does not have to make sure of the $\sigma_y$ rate, he can check the bit flip rate and phase flip
rate $after$ the purification. If initially the error rate is indeed symmetrically distributed and 
the total flipping rate is less than $50\%$, he must be able to distill highly pure entangled state finally. This means, with the real 
entanglement purification, a 4-state protocol can tolerate a flipping rate of $33.3\%$ $if$ the 
channel noise is symmetric. (One does not have to test whether the channel noise is symmetric in the protocol, Alice and Bob just go 
ahead to make the purification
even though the rate of  $\sigma_x$ and the rate of  $\sigma_z$ 
 are larger than $25\%$. If the channel $is$ symmetric, he will be able to verify that both bit flip rate and phase flip rate are
0 after the purification).
In our entanglement assisted protocol, the tolerable bit flip rate or phase flip rate for the 6-state protocol is also raised to $30\%$. 
Our entanglement
assisted protocol can tolerate the highest error rate among all proven 
secure prepare-and-measure protocols raised so far for both  4-state case and 6-state case.

Practically speaking, a quantum channel is normally noisy therefore the raw 
key string with Bob and the string with Alice are normally not identical. 
Moreover, it is possible that, in the case Alice and Bob try to make the quantum key distribution,
the channel noise could actually come from Eve. In such a case the raw key 
can be significantly correlated with Eve's quantum state. 
Even in the case that no bit flip or phase flip is detected on the subset of the check bits, the raw key
is still not completely reliable or secure  because there could be still a few bit flips or phase flips for 
the code bits which are unchecked.
To obtain the highly reliable and secure key, one has to take the error
correction (EC) and privacy amplification (PA) to the raw key and then
use the final key which should be unconditionally secure and perfectly reliable.
Although the classical EC can be used to remove all bit-value errors (with a high probability) therefore to help Alice and Bob obtain
 a reliable key, it's not  transparent that whether the classical PA can 
really work here: Eve may first store her qubits which are correlated with the raw key. After Alice and Bob 
complete the EC and PA, 
 she then takes an optimized measurement to her qubits directly
attacking the final key.  
A strict mathematical proof for the unconditional security 
is non-trivial\cite{mayersqkd,others,others2}.
The security proof of QKD is greatly simplified if one connects this with
 the quantum entanglement purification protocol (EPP)\cite{BDSW,bdswa,deutsch}.
The main idea is conceptually 
simple and clear: Alice and Bob first share a number of
raw  entanglement pairs and then purify them to almost maximally entangled
pairs and measure each of them to obtain the final key\cite{deutsch}.
The  strict mathematical security proof with EPP was given by Lo and 
Chau\cite{qkd}.   
Interestingly, it was then shown by
Shor and Preskill \cite{shorpre}  that Lo-Chau entanglement purification
based protocol 
can be reduced to the quantum error correction (QEC) protocol with 
one-way communication and the QEC protocol is equivalent to BB84 protocol
followed by the  classical EC and PA done by decoding a classical  CSS code.
 Shor-Preskill protocol\cite{shorpre} works as long as the measured bit flip error and phase flip error rates are 
less than 11\%, the point at which the Shannon rate hits 0.
Note that this threshold is lower than that of certain EPP protocols with 
two-way communications (2-EPP): the 2-EPP with error rejection  works 
as long as the  summation of measured bit
error and flip error rates are
less than a 50\%.  In such a case,
Alice and Bob randomly choose two pairs and 
measure the parity of 2 qubits in each side and discard both pairs if the parities disagree and keep one pair and discard the other 
pair if the parities agree with each other. In such a way, the bit flip error is reduced
if they measure the parity in $Z$ basis; the phase flip error is reduced
if they measure the parity in $X$ basis.    X, Y and Z  represents the basis in the eigenstates 
of operator $\sigma_x,\sigma_y$ and $\sigma_z$ respectively. The vector  representation for the two level state is
$|0\rangle=\left(\begin{array}{c}1\\0\end{array}\right); |1\rangle=\left(\begin{array}{c}0\\1\end{array}\right)$.
This 2-EPP with error rejection done in both Z basis and X basis cannot be  
reduced to a classical protocol. In a classical protocol, Alice just sends
Bob the  qubits prepared in X or Z (or Y) basis randomly and then they carry out EPP task as if they were sharing 
a number of raw entangled pairs. Or equivalently,  in a classicalized 
protocol, Alice had measured her halves of the entangled pairs before the protocol was started. They can still do the parity
 comparison on Z basis, but they 
will not be able to do so in X basis: they are never able to know 
what the parity values should be if
they had really used entangled pairs here therefore they don't know whether
they should discard both bits or keep one bit. 
 Very recently, motivated for  higher bit error rate tolerance
and  higher  efficiency, Gottesman and Lo\cite{gl} 
studied the classicalization of 2-EPP. It has been shown there that
a 2-EPP can be classicalized iff their action after the phase parity
comparison is deterministic. In such a case they can carry out the task as if
they were doing the 2-EPP on entangled pairs.
Based on this observation, a new QKD protocol was given there\cite{gl} with
partially two way communications: in removing the bit flip errors, the
error-rejection method is used with two way communication; in removing the
phase flip error, the error $correction$ method is used with one way communication, 
i.e., Alice asked Bob to measure the syndrome of certain randomly chosen 3 qubits
and Bob will use the majority rule to decide whether to take a phase flip
operation to one  of the 3 qubits. 
 This method has increased the 
tolerable bit error rate of noisy channel to 18.9$\%$ and 26.4$\%$ 
for 4-state QKD and 
6-state QKD, respectively. Very recently, these values have been upgraded
to 20$\%$ and 27.4$\%$ by Chau\cite{chau}.

In this paper, we propose a revised scheme which is also uncinditionally secure and which can further increase these 
thresholds on bit error rates given the independent channel errors.
We propose to let Alice send Bob the quantum 
states randomly chosen from $\{\frac{1}{\sqrt 2}(|00\rangle + |11\rangle),
\frac{1}{\sqrt 2}(|00\rangle - |11\rangle), 
|00\rangle, |11\rangle\}$. As we shall see, these states are just the quantum
phase-flip
error-rejection (QPFER) code for the BB84 state $\{|0\rangle, |1\rangle,
\frac{1}{\sqrt 2}(|0\rangle + |1\rangle),\frac{1}{\sqrt 2}(|0\rangle - |1\rangle)\}$. When Bob receives them, 
he first decodes each two-qubit code
to the one-qubit state (a BB84 state) 
and then carry out the rest tasks
of EC and PA.
In decoding, Bob discards those codes which contains one bit-flip
 error therefore
after decoding the phase-flip error for the accepted bits are greatly decreased. 
As we are going to show, we may take this advantage to increase the threshold of the tolerable bit error rate caused by the s-channel. 
The advantage of a higher tolerable error rate
is limitted to the cases where the 2-qubit QPFER code works, e.g., in the case that the channel noise is uncorrelated.
 But security part of our  protocol is unconditional, i.e., it is  secure under whatever 
types of attack. In other words, Alice and Bob don't care about the actual noise pattern (correlated or uncorrelated) 
in the QKD, they just test the error rate after decoding and then go ahead to distill the final key by our protocol.
The final key is always secure no matter what type of noise has been actually happened to the raw qubits.
 Subtle points for the unconditional security with conditional  advantage of the protocol is given in the
following section. In particular, the protocol is totally different from
the almost useless protocol which is only secure with uncorrelated channel noise.  
\section{Subtlities of unconditional security with conditional advantage of our protocol}
To a QKD protocol, one may evaluate it by  several criterion, e.g., the 
security, the efficiency and the key rate.
The unconditional security is the first important thing. Without this
prpoerty, the protocol is almost useless even if it has other advantages
such as a high key rate. After all, the reason we use $quantum$ key distribution is
due to the temptation of its {\it unconditional security}. In this logic,
a $conditional$ secure QKD protocol with unverifiable condition for
security is useless, because it is actually insecure and has lost
the assumed advantage to a classical key distribution. 
For example if a new protocol is proposed with advantages in efficiency
or economy, but the security is only proved based on the assumption
of individual attack, this new protocol is useless unless the proof of
unconditional security is completed, i.e., the proof of the security
under whatever type of attack.

An unconditionally secure QKD protocol means that Eve's information to 
the final key is exponentially close to 0. To Alice and Bob, if they carry out 
that protocol, the worst result there is that Alice and Bob may get 0 bit for the 
final key, i.e., when the channel noise 
(including the disturbance caused by Eve's attack) is higher than the threshold of
the protocol itself. The ``efficiency'' is defined by the threshold of the channel
noise for the protocol. If the channel noise is larger than that, Alice and Bob will
obtain zero bit for their final key. 
In almost all proven unconditionally secure protocols so far, the efficiency is dependent on
the channel error rate, but it is independent
of the channel error pattern itself, i.e., no matter it is a coherent error or uncorrelated error,
the threshold is the same. These protocols are unconditional secure  with threshold value of
noise which is universal to all noise patterns, correlated or uncorrelated. 
Here in our protocol,  the threshold is dependent on the 
channel error pattern while the security is independent of the error pattern (which is required by
the term ``unconditional security''). Note that our protocol is totally different from the ones with
a conditional security. Our protocol is unconditionally secure with a conditional advantage in noise
threshold. 
 
Security proof of a QKD protocol is normally related to the channel noise. Here we shall
use the term ``s-channel'' for the physical channel which is supposed to transmit the 
qubits from Alice
to Bob. This s-channel can be an optical fiber, or can be just the nature. We shall use the
term  ``r-channel'' for the actual channel in the QKD, the noise of ``r-channel'' is supposed to 
be under control of Eve. in the security proof. In particular, if there is no Eve., r-channel is 
just s-channel. 
Suppose Alice and Bob decide to do QKD through a certain s-channel (e.g., optical fiber F)  by 
using Gottesman-Lo protocol\cite{gl,chau}.
 Before they do
so they have been very sure of  the propeties of their s-channel noise. They find the channel
noise is all independent to each qubit and the bit error rate of that s-channel is $21\%$. 
(It is to many people's intution that the errors should be independent if the different qubits are spatially
separated sufficiently. Also we can choose to use the type of fiber which only
causes independent errors to qubits.) In such a case,
if we are limitted to the previously proposed 4-state  prepare-and-measure protocols, we can do nothing for the quantum
key distribution because the error rate is larger than the thresholds of all those protocols.
However, in our protocol, Alice will first encode each BB84 qubit by a 2-qubit quantum error-rejection code 
and then send each quantum code to Bob. Bob will first take the parity check and decoding after ha receives the codes.
He will test the error rate and distill the final key with Alice $after$ decoding.
Due to the fault tolerance property of the error rejection code, the error rate of the raw bits after decoding
is expected to be reduced and the net effect here is that our 4-state protocol can tolerate an error rate up to $26\%$ for 
the s-channel.
In the supposed case of $21\%$ error rate for the symmetric s-channel where all errors are independent, 
our protocol will still work.

The advantage of a higher error threshold is not an unconditional advantage. It is only for the cases 
where the 2-qubit error rejection code indeed works. One example is that the channel only causes uncorrelated
errors. However, the security of our protocol is unconditional. That is to say, no matter what type of attack Eve.
may use, the final key obtained by our protocol is always secure.
In a real QKD process the qubits
could be transmitted through Eve. or
 the property of s-channel could be changed due to Eve's attack therefore the 2-qubit error-rejection code does not 
work properly as what is expected. In such a case, after decoding, Alice and Bob will find that the error rate is 
unexpectedly high and
they will choose to either abort all the raw bits or continue the rest steps of the protocol and obtain a shorter
final key to which Eve.'s information is exponentially close to 0.
But in cases there is no Eve., or Eve.'s attack is insignificant, the 2-bit error-rejection code works perfectly
or almost perfectly, the advantage of our protocol appears: one can obtain the final key 
in the case  that the s-channel noise is higher than the thresholds of all previously known secure 
prepare-and-measure protocols.  

If Eve always disturbs the qubits very sigficantly, our protocol
will not work at all even though the error rate of s-channel is lower than the threshold.
 However, in such a case no protocol will work.  
A quantitative comparison of different protocols on 
Eve's cost to destroy a protocol or to shorten the final key is not a topic
of this work. 
\section{ Gottesman-Lo protocol:  error-rejection and error-correction } 
\begin{figure}
\epsffile{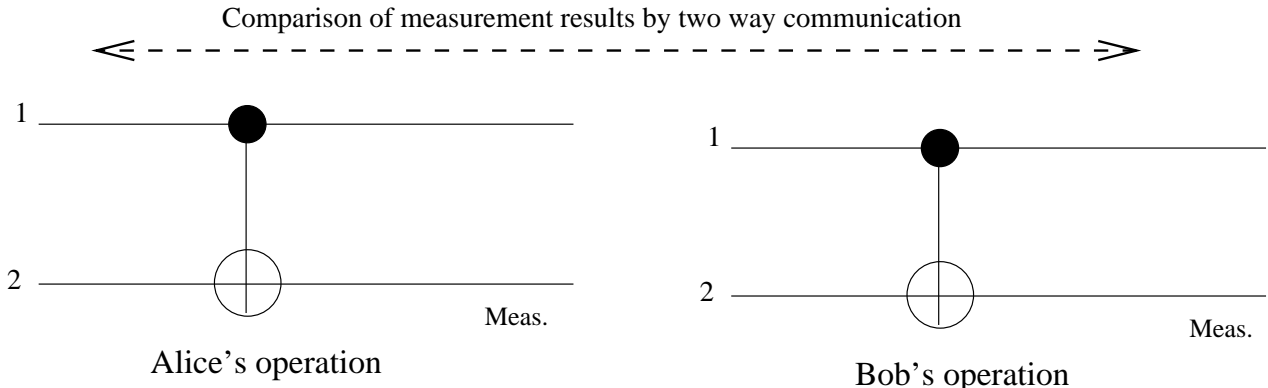}
\caption{ Controlled-not (C-NOT) operation used for the  bit-flip error-rejection. The horizontal lines marked by 1 and 2 are pair $1$ and $2$ respectively.
 Alice and Bob  compare the measurement outcomes of the target qubit $2$.
If they choose Z basis for C-NOT operation and the measurement, it is 
equivalent to measure $Z_1Z_2$; if they choose to use X basis, it is 
equivalent to measure $X_1X_2$ on each side.}\label{cnot0} 
\end{figure}

There are many ways to do entanglement purification. However, not all of them can be used 
for the security proof of a QKD protocol without using quantum storage. 
To carry out such a task
one must first study an entanglement purification based QKD protocol and then show that the 
protocol is classicalizable, i.e.,
to show that it is equivalent to the case where  
Alice measures all her halves of entangled pairs initially and continue with all other steps in the protocol.
Different types of EPP may tolerate  different flipping rates of the channel.    
We now first analyse the reason why
the currently existing protocols\cite{gl,chau} does not reach the theoretically allowed threshold of the channel flipping rate. To prove a secure
QKD protocol, one may first consider an entanglement purification protocol and then find out the corresponding
QKD protocol based on that.
In general, one has two simple ways to purify entangled  pairs shared
by spatially separated parties, Alice and Bob.
One method is the error-$rejection$\cite{BDSW}: The raw pairs are randomly divided into  
many 2-pair groups.   
To each group, Alice and Bob  measure the parity on each side. 
If they obtain the same parity value, they discard the target pair and keep the control pair (see Fig. \ref{cnot0}). If the
parity values are different, they discard both pairs. In doing the parity check, they can choose
the measurement basis of $Z_1 Z_2$ to reduce the bit flip errors or the basis of $X_1 X_2$
to reduce the phase flip errors.
 Suppose the initially shared raw pairs between Alice and Bob bear the $\sigma_x,\sigma_y,\sigma_z$ errors 
are $p_{x},p_{y},p_{z}$, respectively. 
Let $p_{I}=1-p_{x}-p_{y}-p_{z}$.
After one round of bit-flip error-rejection, the error rate for the survived 
pairs is changing by the following
iteration formula as given by Chau\cite{chau}:
 \begin{equation}
  \left\{ \begin{array}{rcl} p^{EP}_I & = & \displaystyle\frac{p_I^2 +
   p_z^2}{(p_I + p_z)^2 + (p_x + p_y)^2} , \\ \\
   p^{EP}_x & = & \displaystyle\frac{p_x^2 + p_y^2}{(p_I + p_z)^2 +
   (p_x + p_y)^2} , \\ \\
   p^{EP}_y & = & \displaystyle\frac{2p_x p_y}{(p_I + p_z)^2 + (p_x +
   p_y)^2} , \\ \\
   p^{EP}_z & = & \displaystyle\frac{2p_I p_z}{(p_I + p_z)^2 + (p_x +
   p_y)^2} .
  \end{array}
  \right. \label{E:LOCC2EP_map_asy}
 \end{equation}
Exchange $p_x$ and $p_z$ above we can obtain the iteration formula by phase
 error-rejection operation.
 By taking the error-rejection operation alternately in $Z$ basis and $X$ basis for many rounds,
one can always distill out maximally entangled pairs asymptotically from the raw state
$\rho$ if $\langle \Phi^+|\rho |\Phi^+\rangle > 1/2$, where
$|\Phi^+\rangle=\frac{1}{\sqrt 2}(|00\rangle + |11\rangle)$.
This is to say, with the error-rejection method,   
the theoretical upper bound of tolerable channel flipping rate
for QKD distribution can be reached. For  the symmetric channel,
the bit-flip rate or the phase-flip rate can be as high as $33.3\%$
no matter whether we use the 4-state protocol or 6-state protocol. 
However, the error-rejection
method above cannot be clasicalized to the prepare-and-measure QKD protocol since if Alice had measured
her halves in Z basis initially, Alice and Bob would have no way to take phase-flip error-rejection.
Note that the error-rejection method means they have to decide whether they
should keep one bit or discard both bits according to the measurement result.

\begin{figure}
\epsffile{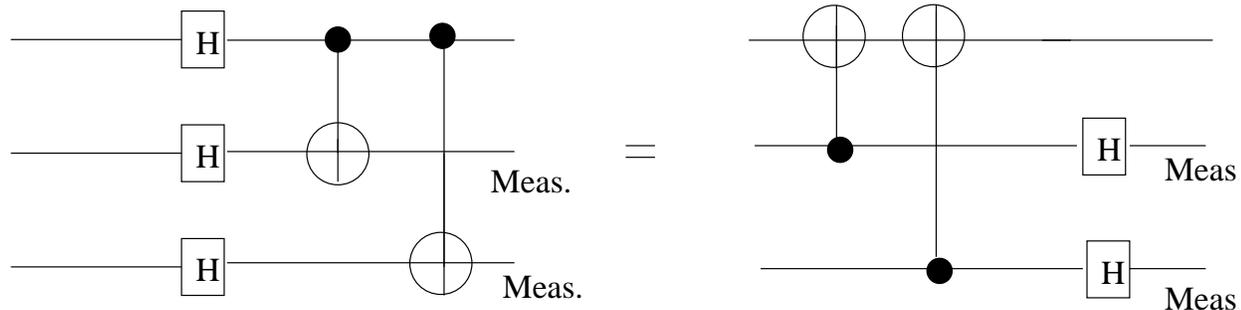}
\caption{Two equivalent ways to measure the operators $X_1X_2$ and $X_1X_3$. This figure is taken from the paper by Gottesman and Lo.}
\label{pstep}
\end{figure}

To overcome this barrier, Gottesman and Lo proposed to use the  error-$correction$ method to reduce the phase-flip error: 
they $correct$ a possible error by $[3,1,3]_2$ code instead of discarding
the the corrupted pairs, once a possible error is detected. 
To do so, Alice and Bob randomly choose 3 pairs in one group. They each measure
the parity of qubits 2,3 and qubit 1,2 and compare both 
values (Fig. \ref{pstep}).
After the comparison, they  decide whether to take a flip operation to pair 1. They keep pair 1 and discard   
pair  2 and pair 3. With a high probability that pair one is now  free of phase flip error if the parity checks   are done in $X$
basis.  However the tolerable initial error rate is decreased with this method towards the distillation of maximally
entangled pairs.
{\it This implies that the error-rejection can be more effective than error-correction in the entanglement distillation.}

In Gottesman-Lo QKD protocol, Alice and Bob distill the classical
 bits as if they were distilling the entangled pairs: no one knows whether Alice had measured
her halves of the entangled pairs initially. Therefore the security of the protocol is equivalent to
that in a real entanglement distillation protocol, which has been proved to be unconditionally
secure\cite{qkd}.

Gottesman-Lo protocol reduces the bit flip error by error-rejection and reduces the phase flip error
by error-correction.
 Although in the case of real entanglement distillation, Alice may often
need to phase flip certain qubit according to the parity comparison result
for phase-flip error-correction in the entanglement purification protocol.
However, if the final purpose is to set up the faithful and secure
key only, Alice need not really take any phase flip. Instead, she may simply use the parity of randomly chosen 3 bits
as the new bit value after ``error-correction'' (see Fig.\ref{pstep}). 
 
The phase flip does not affect the final bit result therefore omitting
the phase flip will not affect the reliability of the final key. That is,
the final key is as faithful as that in the case Alice takes phase flip to 
her qubits as required by the standard entanglement purification. 
Moreover, one can also find that omitting the phase flip operation does not 
affect the security either. 
Consider the case that Alice never takes phase flip to her qubit but
she  keeps the information in the mind that
which qubit should be phase flipped and then continue  the distillation, 
they will finally obtain
$|\chi_1\rangle\otimes|\chi_2\rangle\cdots |\chi_n\rangle$. Each $|\chi_i\rangle$ is either $|\Phi^+\rangle$
or $|\Phi^-\rangle=\frac{1}{\sqrt 2}(|00\rangle-|11\rangle)$ and Alice explicitly knows the state of each pair (with a probability exponentially close to 1).
That is to say, if Alice never takes any phase flip, Alice and Bob will  share a  product of different maximally
entangled state. Note that the shared pairs are in a pure state. Therefore 
those shared pairs, with a probability exponentially close to 1, are unentangled with any third party.
The security in such a case is totally equivalent to the case where all shared pairs are in state $|\Phi^+\rangle$.
Further, even Alice never keeps any information in her mind on which qubit should be phase flipped, the security is unchanged. Because even in such a case,
the finally distilled pairs are also unentangled with any third party.
The only thing that is important here is that Alice would be able to know
the state of finally distilled pairs if she had remembered the information
 on which qubit should be phase flipped.
 
One may further reduce the above protocol.
Taking the fact that the phase flip operation can actually be ignored, Alice can then choose to measure each of her halves 
of EPR pair in $Z$ 
basis initially and send the other halves to Bob and then carry out the same entanglement purification scheme as if 
they were sharing the entangled pairs. This reduction should not affect the reliability of Bob's final key because
it does not affect any bit value  in $Z$ basis. Note that the operation commutes with all operations taken by Alice latter on.
This reduction should be as secure as before because Eve will never know
whether Alice had measured her halves of EPR pairs in the beginning, even with the full collaboration with Bob.
Suppose the reduction at this step is insecure. We have already known that  the final key is secure
if Alice had not measured her halves of pairs . Then  we shall get the confliction that
Eve. and Bob  will be able to know whether Alice had measured
her qubits in the beginning by checking whether Bob's final key is significantly correlated with Eve's assumed key.

From the above analysis we spot one important fact in Gottesman-Lo protocol\cite{gl} and its modified 
version\cite{chau}:
Alice and Bob are  only allowed to take error-$correction$ but  not allowed to take error-$rejection$ operation
to treat the phase flip error, otherwise it cannot be classicalized.
But error-$rejection$ operation can be more effective in removing errors. 
This is the reason why the error rate threshold cannot reach the theoretically allowed value, i.e., the threshold
for the real entanglement distillation with all error-rejection method.

\section{ Phase-flip error rejection with 2-qubit error rejection code.}
To improve the tolerable channel error rate threshold in quantum key distribution, one may naturally consider the possibility
of using error-rejection method to reduce the phase error also. Naively, one may consider the real 2-EPP. But that requires
the quantum storage which is obviously impractical  with our current technology. However, with the quantum error-rejection
code as we are going into, one can take the error-rejection $quantumly$ therefore
any quantum storage is unnecessary. However, if we want to use quantum phase error-rejection code to remove the phase flip
error in all rounds, we must then use a large concatenated quantum code which is also impractical. Keeping this point 
in mind, we choose to use quantum error-rejection only at the first round therefore we only need a 2-bit quantum code to 
encode each initial qubits with Alice:
\begin{eqnarray}\nonumber
|0\rangle|0\rangle\longrightarrow (|00\rangle + |11\rangle)/\sqrt 2\\
|1\rangle|0\rangle\longrightarrow (|00\rangle - |11\rangle)/\sqrt 2.
\label{pfer}\end{eqnarray}
Here the second $|0\rangle$ state qubit in the left hand side of the arrow is
the ancilla for the encoding.   
This code is not assumed to reduce the errors to qubits in all cases.
But in the case that the channel noise is uncorrelated
or nearly uncorrelated, it works effectively. We shall consider the ideal
case that the channel errors are uncorrelated here.
Therefore the initial perfect EPR pair $|\Phi^+\rangle$ with Alice is encoded by the following formula before sending half of it
Bob over noisy channel:
\begin{eqnarray}
E[|\phi^+\rangle]=\frac{1}{2}\left[|0\rangle_A(|0\rangle_{B1}|0\rangle_{B2}+|1\rangle_{B1}|1\rangle_{B2})
+|1\rangle_A(|0\rangle_{B1}|0\rangle_{B2}-|1\rangle_{B1}|1\rangle_{B2})
\right]\label{epair}
\end{eqnarray} 
Alice then sends qubits B1 and B2 to Bob. In receiving them, Bob first takes a parity check, i.e. measures $Z_1Z_2$.
Note that this measurement does not destroy the code state itself. Moreover, any linear superposition of 
$|00\rangle$ and $|11\rangle$ is an eigenstate of $Z_1Z_2$ with eigenvalue 0;  any linear superposition of 
$|01\rangle$ and $|10\rangle$ is an eigenstate of $Z_1Z_2$ with eigenvalue 1.
If he obtains 1, Alice and Bob discards the code, if he obtains 0, Bob decodes the code:  he takes  
a measurement in $X$ basis to one qubit (say, $B1$), if he obtains $|+\rangle$, he takes a Hadamard transformation $H=\frac{1}{\sqrt 2}\left(\begin{array}{cc}
1 & 1\\1 & -1\end{array}\right)$ to the other qubit (B2);
if he obtains $|-\rangle$ for qubit B1, he takes the Hadamard transformation to qubit B2 and then flips qubit B2 in Z basis. 
In such a way, Alice and Bob may share the raw pairs of qubit A and B2.
The phase flip error to the survived raw pairs will be decreased.
Suppose the channel error rates of $\sigma_x,\sigma_y,\sigma_z$ types  
 are $p_{x0},p_{y0},p_{z0}$, respectively. Let $p_{I0}=1-p_{x0}-p_{y0}-p_{z0}$.
With a probability of $p_{I0}^2$ there is no error to both qubit B1 and B2.
In such a case, the state after decoding is exactly $|\Phi^+\rangle$. Explicitly, with no flip happens
the total state for $A$, $B1$ and $B2$ before decoding is
\begin{eqnarray}
E|\Phi^+\rangle = |0\rangle_A(|+\rangle_{B1}|+\rangle_{B2}+|-\rangle_{B1}|-\rangle_{B2})+
 |1\rangle_A(|+\rangle_{B1}|-\rangle_{B2}+|-\rangle_{B1}|+\rangle_{B2}).
\end{eqnarray}
Note that the total state is symmetric to qubit B1 and B2.
Suppose Bob takes measurement in X basis to qubit $B1$ for the decoding.  From the formula above we can see that the state 
for qubits A and B2 will 
be collapsed
to $|0\rangle|+\rangle+|1\rangle|-\rangle$ if he obtains $|+\rangle$ for qubit B1, after a Hadamard transformation, the shared pair 
is changed to $|\Phi^+\rangle$ for sure. Qubits A and B2 will be collapsed to
 $|0\rangle|+\rangle-|1\rangle|+\rangle$ if he obtains $|-\rangle$ for qubit B1. In such a case, he will first take a Hadamard transformation to B2 and then flip it in Z basis.  After these operations, the shared pair
is also changed back to state $|\Phi^+\rangle$ for sure.
If one of the transmitted qubit bears a bit flip error (including both $\sigma_x$ and $\sigma_y$ error) while the other
transmitted qubit does not bear the bit flip error (e.g., it can be of no error or  of $\sigma_z$ error),
the code will be discarded for sure after the parity check. The probability for this type of event is
$2(p_{x0}+p_{y0})(p_{I0}+p_{z0})$. If both transmitted qubits bear bit flip errors with arbitrary phase flip errors, or
none of them suffer a bit flip error with arbitrary phase flips, the 2-qubit code will pass the parity
check and be accepted. In such a case Alice and Bob will share a wrong state after  decoding. However, the probability for such cases
are small. We list the probability distribution for all possible states after decoding in the following table:
\begin{center}
\begin{tabular}{rrrr}\hline
joint channel  error  \vline & probability \vline& state after decoding \vline&  
raw pair error  
\\ 
\hline 
$I\otimes I$  \vline& $p_{I0}^2$  \vline& $|\phi^+\rangle$   \vline& $I$\\
\hline
$\{I\otimes \sigma_z\}$  \vline& $2p_{I0}p_{z0}$  \vline& $|\psi^+\rangle$  \vline& $\sigma_x$
\\
\hline 
$\sigma_z\otimes \sigma_z$  \vline& $p_{z0}^2$  \vline& $|\phi^+\rangle$   \vline& $I$\\
\hline
$\{\sigma_y\otimes \sigma_y\}$  \vline& $p_{y0}^2$  \vline& $|\phi^-\rangle$  \vline& $\sigma_z$
\\
\hline 
$\{\sigma_x\otimes \sigma_x\}$  \vline& $p_{x0}^2$  \vline& $|\phi^-\rangle$  \vline& $\sigma_z$
\\
\hline
$\{\sigma_x\otimes \sigma_y\}$  \vline& $2p_{x0}p_{y0}$  \vline& $|\psi^-\rangle$  \vline& $\sigma_y$
\end{tabular}
\end{center} 
We use $\{\alpha\otimes\beta\}$ to denote both $\alpha\otimes\beta$ and $\beta\otimes\alpha$ in the most left column above.
There are other types of joint errors to the two transmitted qubits besides those listed in the above table. However,
all codes with those types of unlisted joint channel errors will be discarded after Bob's parity check operation. After renormalize 
the probability distribution for
each term in the above table we can obtain the error rate distribution for the survived raw pairs shared by Alice and Bob. 
 \begin{equation}
  \left\{ \begin{array}{rcl} p_I & = & \displaystyle\frac{p_{I0}^2 +
   p_{z0}^2}{(p_{I0} + p_{z0})^2 + (p_{x0} + p_{y0})^2} , \\ \\
   p_z & = & \displaystyle\frac{p_{x0}^2 + p_{y0}^2}{(p_{I0} + p_{z0})^2 +
   (p_{x0} + p_{y0})^2} , \\ \\
   p_y & = & \displaystyle\frac{2p_{x0} p_{y0}}{(p_{I0} + p_{z0})^2 + (p_{x0} +
   p_{y0})^2} , \\ \\
   p_x & = & \displaystyle\frac{2p_{I0} p_{z0}}{(p_{I0} + p_{z0})^2 + (p_{x0} +
   p_{y0})^2} .
  \end{array}
  \right. \label{errorrate}
 \end{equation}
With this formula, the phase flip error to the shared raw pairs is  obviously reduced. The formula is similar to the error-rejection formula
by directly measuring the parity of two pairs on each side. Therefore such a code can be  more effective
than $[3,1,3]_2$ error correction code in reducing phase-flip errors and can cause advantages
to the QKD protocols.
 Note that it is a bit subtle that the phase error to the decoded state is caused
by bit-flip channel errors.
\section{Entanglement assisted protocol.}
We first consider the following entanglement purification based protocol with quantum storages:
\begin{itemize}
\item[\bf 0] Before they carry out the protocol, Alice and Bob test
there channels by the error rejection code in above section. They find
Eq.(\ref{errorrate}) alomost always (approximately) holds. 
\item[\bf 1]
Alice prepare $N$ copies of EPR state $|\Phi^+\rangle$,
$|\Phi^+\rangle=\frac{1}{\sqrt 2}(|0\rangle_A|0\rangle_{B1}+|1\rangle_A|1\rangle_{B1})$. She also prepares N ancilla qubits which are all in state 
$|0\rangle$.  
We shall use the subscript B2 for each of theses N ancilla qubits. To each pair, she will keep qubit A with herself.
\item[\bf 2]
Alice takes unitary transformation  given in Eq.(\ref{pfer}) to qubits B1 and 
B2. B1 and B2 are now the phase-flip quantum error-rejection code for the
original state of qubit B1. Now the total state of qubit A, B1, B2 is in the form
of eq.(\ref{epair}).
\item[\bf 3]
Alice sends each 2-qubit code (B1,B2) to Bob and keeps 
qubits A. 
\item[\bf 4]
Bob takes parity check to each 2-bit code 
 in basis of $|0\rangle\langle 0|, |1\rangle\langle 1|$. 
That is, Bob measures $Z_{B1}Z_{B2}$.
He discards both of them  whenever he obtains 1 and stores both qubits 
whenever he obtains 0.
\item[\bf 5]Bob decodes each of the survived 2-bit code:
To each code that has passed the parity check, he measures one qubit (e.g., qubit B1) in X basis.  
If he obtains $|+\rangle$, 
he takes a Hadamard transformation 
to the other qubit (qubit B2)
and saves it as the decoded qubit; if he obtains $|-\rangle$ as the measurement result for B1, he will first take a 
Hadamard transformation to B2 and then
flip B2 in Z basis and save it as the decoded qubit. 
\item[\bf 6] Bob announces which codes have passed the parity check and have been decoded. Alice discards those 
qubits with her which were entangled with
Bob's discarded codes. Now Alice and Bob share a number of raw entangled pairs
with each of them containg qubit A and qubit B2.
Alice then randomly chooses a number of her qubits and measure each of them in 
either Z basis or X basis. She announces which qubits have been chosen and the measurement basis and results of each of them. 
According to her announcement,
Bob measures his halves of those raw pairs in the same basis as chosen 
by Alice.
He compares his measurement results with those announced by Alice. 
 If too many of them disagree, they abort
the game and then restart from the beginning. 
If the error rate is acceptable,
they continue the game with the following steps\cite{sremark}. 
\item [\bf 7] Alice and Bob then take the bit-flip error-rejection\cite{gl}: they divide the rest of the survived 
raw pairs into many 2-pair groups. The two pairs for each individual groups are randomly chosen.
They measure the parity of each group in Z basis at each side  and compare the results. They discard those groups whose
parities on two sides disagree; they discard the target pair and keep the 
control pair (see Fig.\ref{cnot0})if the results at two sides 
agree with each other.
They can repeatedly take this bit-flip error correction for $g$ round.  
\item[\bf 8] They can then take the $[r,1,r]_2$ phase-flip error-correction code as proposed in \cite{chau}
to reduce the phase-flip error. They should choose the appropriate $g$ and $r$ value so that both bit-flip
error and phase-flip error are small enough therefore the shared pairs can be distilled to maximally
entangled by a certain classicalizable distillation method in the next step. 
\item[\bf 9] 
They further purify the survived pairs by a classicalizable distillation method
so that the error rate
of their shared pairs are exponentially small.  The classicalizable distillation method
means the method which can be reduced
to an equivalent prepare-and measure scheme\cite{gl} where no quantum storage is required.
\end{itemize}
Note that this scheme requires the quantum storage. We shall classicalize it to an equivalent
form latter therefore a
quantum storage is unnecessary. For the moment we first 
 calculate the largest tolerable channel error rate of the protocol.

The two-qubit error-rejection code decreases the phase flip error and increases
the bit flip error of the shared raw pairs. 
After the quantum decoding Alice and Bob will reduce the bit flip error rate
by bit flip error-rejection.
 The error rate for the survived pairs after one round bit flip error-rejection is given by  Eq.(1).
This is the formula which is used iteratively  in the multi-rounds bit-flip error-rejection. 
In doing the error rejection for bit-flip error, the phase flip rate and also perhaps the total error rate
is increased. 
After some rounds of bit flip error rejection, one must then reduce to the phase flip error rate by the
scheme of phase flip error-correction.
In the original Gottesman-Lo protocol,  $[3,1,3]_2$ code is used.
Actually, any $[r,1,r]_2$ error correction code can be used for that task and any  $[r,1,r]_2$ can be 
classicalized\cite{chau}. In our protocol, we have followed Chau's protocol: carry out the bit flip error-rejection
repeatedly and use an appropriate  $[r,1,r]_2$ code to reduce the phase error.
After the phase error correction, the new error rates satisfy the following inequalities provided that $p_I > 1/2$.
\begin{equation}
  \left\{ \begin{array}{rcl} p^{PEC}_x + p^{PEC}_y & \leq & r
   (p_x + p_y) ,  \\
   p^{PEC}_y + p^{PEC}_z & \leq & \left[ 4 (p_I + p_z) (p_x +
   p_y) \right]^{r/2} 
    \leq  e^{-2 r (0.5 - p_z - p_y)^2} ,
  \end{array}
  \right. \label{E:PEC_map_asy}
 \end{equation}
This shows that, given $p_{x0},p_{y0},p_{z0}$, 
if there exits a finite number $k$, after $k$ rounds of bit-flip error-rejection,
we can find a $r$ which satisfy
\begin{eqnarray}\nonumber
r(p_x+p_y)\le 5\%\\
e^{-2r(0.5-p_z-p_y)^2}\le 5\%.  
\end{eqnarray}
One can then obtain the unconditional secure and faithful final key with a further purification through any
classicalizable methods.
 
To the above 4-state protocol, even though 
$p_{y0}$ value is not detected, we don't have to assume $p_{y0}=0$. 
We actually need not to know it.
What we need to know is each types of error rates to the {\it raw pairs}, i.e. $p_x,p_y,p_z$.  Based on the
information of $p_x,p_y,p_z$ we then decide whether the shared raw pairs are distillable by the remained steps in the
scheme. In other words, this is equivalent to the case where Alice directly sends the EPR halves to Bob through
an un-symmetric noisy channel with flip rates of $p_x,p_y,p_z$ instead of $p_{x0},p_{y0},p_{z0}$.
That is to say, the role of the quantum phase flip error-rejection code is to replace the natural channel error rate
$p_{x0},p_{y0},p_{z0}$ by the effective channel error rate of   $p_x,p_y,p_z$ which are linked by eq.(\ref{errorrate}).
Note that a quantum $bit$-flip error-rejection code will not help to improve the tolerable error rate of Gottesman-Lo protocol\cite{gl}
because the effect of that is equivalent to that of the classical bit-flip error-rejection as used there.
In the 4-state protocol, we don't detect the $\sigma_y$ error therefore
we have to assume $p_y=0$ after the quantum parity check and decoding.
This shows that, if the channel noise is symmetric, after the quantum decoding, both $\sigma_z$ error and $\sigma_y$
error are  reduced, i.e., the detectable phase error rate( including both $\sigma_z$ and $\sigma_y$ error rate) 
has been   reduced in a rate as it should be in the case of 
symmetric noisy channel. We then start from the un-symmetric error rate with assumption $p_y=0$ and $ p_x,p_z$ being the
detected bit-flip rate and phase-flip rate, respectively. 
After the calculation, we find that the tolerable error rate of bit flip or phase flip is $26\%$ for the 4-state protocol.
 Moreover, in the case that the channel error itself is un-symmetric, e.g., $p_{y0}=0;p_{x0}=p_{z0}$,
the tolerable channel error rate for our protocol  is   $p_{x0}=p_{z0}\le 21.7\%$.
 \section{The protocol without quantum storage}
The above QKD scheme requires a quantum storage for both Alice and Bob. This is impractical
by our current technology. However, the scheme can be reduced to an equivalent scheme which does not
 require a quantum storage. We now show it in details.

In the protocol above, the phase-flip quantum error-correction code of 
$[r,1,r]$ is used\cite{chau}. This is equivalent to the classical method
of replacing the r bits by one new bit whose value is just the parity of 
those r bits\cite{chau}. For example, one may consider the case of $r=3$. 
As it was shown in Ref.\cite{gl}, the parity  measurement in X basis
can be replaced by 
the equivalent one in Z basis, see Fig.\ref{pstep}. The actual operation is 
just to replace the bit value of qubit 1 by the  parity of 3 qubits.
More generally, the $[3,1,3]_2$ error correction code can be replaced by
$[r,1,r]_2$ code with the majority rule, i.e., replacing the original
one bit value
by the parity of r bits\cite{chau}. 
The initial EPR pairs prepared by Alice will be treated in three different 
ways:
some of them will be used as the check pairs;
some of them will be discarded after the parity check before decoding; 
some of them will be 
used for the entanglement distillation.
To those check pairs, Alice's only operation to qubit A is just a measurement
in either Z basis or X basis, there is no other operation therefore
Alice can choose to measure qubit A before encoding qubit B1.
After Bob's parity check before decoding, some quantum codes will be discarded. The discarded
codes do not affect any results of final key, therefore Alice may choose
to measure qubit A initially in any basis to each of those qubits.
To those qubits which will be used for the distillation,
Alice's all  operations are done in Z 
basis, 
therefore Alice can choose to measure all those qubits A in Z basis before 
encoding B1. 
Moreover, instead of preparing each single qubits and then encoding them
with ancilla by Hadamard transformation and C-NOT gate,
Alice may directly prepare the error rejection code and put down the bit value
corresponding to the code. That is, code $|00\rangle,|11\rangle$, 
$\frac{1}{\sqrt 2}(|00\rangle+ |11\rangle)$ and $\frac{1}{\sqrt 2}(|00\rangle- |11\rangle)$ 
correspond to single qubit 
state $|+\rangle,|-\rangle, |0\rangle,|1\rangle$ respectively.

With the above arguments,  the protocol can be revised to the following protocol $without$ a
quantum storage:
\begin{itemize}
\item[\bf 1]
Alice prepares N 2-qubit quantum codes with each of them
randomly chosen from the set  $\{|00\rangle,|11\rangle, 
\frac{1}{\sqrt 2}(|00\rangle\pm |11\rangle)\}$. Among all of them,
$N/4$ of them are prepared in   $|00\rangle$ or $|11\rangle$ with equal probability and 
$3N/4$ of them are prepared in  $\frac{1}{\sqrt 2}(|00\rangle\pm |11\rangle)$  with equal probability.
She records the the ``preparation basis'' as  X basis for code  $|00\rangle$ or $|11\rangle$;
and as  Z basis for code $\frac{1}{\sqrt 2}(|00\rangle\pm |11\rangle)$\cite{notek}. And she records the bit value
of $0$ for the code $|00\rangle$ or $\frac{1}{\sqrt 2}(|00\rangle + |11\rangle)$; bit value $1$ for the 
code $|11\rangle$ or $\frac{1}{\sqrt 2}(|00\rangle- |11\rangle)$. She sends each  randomly chosen 2-qubit codes to Bob.
\item[\bf 2]
Bob takes parity check to each 2-qubit code in Z basis. 
That is, Bob measures $Z_{B1}Z_{B2}$.
He discards the codes whenever he obtains 1 and he takes the following measurement if he obtains 0 in the parity check:
 he measures one qubit in X basis and the other qubit in either 
X basis or Z basis  with equal probability.
Bob makes a record of his ``measurement basis'' to the decoded qubit
as Z basis ( $\{|0\rangle,|1\rangle\}$) if he measures ``the other qubit''
in basis  X basis ($\{|\pm\rangle\}$); and he records his ``measurement basis''  as X basis if he measures ``the other qubit''
in Z basis\cite{note0}.  
If he obtains $|+\rangle|+\rangle$,  $|+\rangle|0\rangle$, $|-\rangle|-\rangle$ or $|-\rangle|0\rangle$, he records
bit value $0$ for that code; if he obtains  $|-\rangle|+\rangle$, $|-\rangle|1\rangle$, $|+\rangle|-\rangle$ or 
$|+\rangle|1\rangle$, he records
bit value $1$ for that code.
\item[\bf 3] Alice announces her records of ``preparation basis'' of each qubit; Bob compares his measurement basis
with Alice's announced ``preparation basis''. He discards those bit values whose measurement basis disagree with Alice's announcement.
Bob announces which qubits are survived. Bob announces the bit value for those qubits whose ``preparation basis'' and ``measurement basis'' are X.
He also randomly chooses the same number of bits whose ``preparation basis''
and ``measurement basis'' are Z. He  announces their bit values. 
If too many of them disagree with Alice's record,
they abort the protocol.    
\item [\bf 4]
They reduce the bit flip rate by the following way: they randomly group all their unchecked bits with each group
containing 2 bits. They  compare the parity of each group. If the
results are different, they discard both bits. If the results are same, they discard one bit and keep the other.
They repeatedly  do so for a number of rounds until they believe that both bit flip rate and phase flip rate will
be reduced to less than $5\%$ with the next step being taken.
\item[\bf 5] They then randomly group the remained bits with each group containing $r$ bits. They use the parity of each group
as the new bits.
\item[\bf 6] They use any other proven secure classical methods to further reduce the error rate until both bit flip rate and phase flip
rate are negligible, e.g. less than $2^{-50}$. Here ``other classical methods'' includes the Gottesman-Lo method\cite{gl},
the classical CSS code\cite{shorpre}, the concatenated 7-bit code\cite{chau} and so on.  
 \end{itemize}
The above classicalized protocol is totally equivalent to the one based on entanglement purification in the previous section,
therefore it tolerates the same channel error rate as that that of the entanglement purification based protocol. That is to say,
this classicalized 4-state protocol can tolerate a channel flipping rate of $26\%$. Although the quantum storage is removed now,
it still requires the parity check and decoding operation for Bob. We now show how to make it with ordinary linear optics devices.
\section{ Experimental realization.}
\begin{figure}
\epsffile{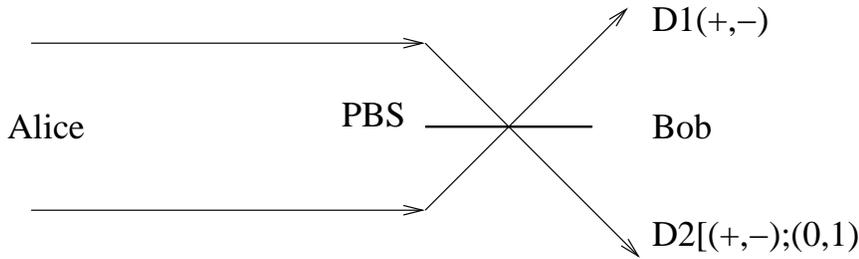}
\caption{ Bob's operation for quantum parity check and decoding. The two horizontal lines are the 2-bit code sent by Alice.
PBS is a polarizing beam splitter which transmits horizontally polarized photons and reflects vertically polarized photons.
D represents for measurement. D1($+,-$) means a measurement in basis $|\pm\rangle$, D2$[(+,-);(0,1)]$ means that Bob takes 
that measurement in either  $\{|\pm\rangle\}$ basis or 
$\{|0\rangle, |1\rangle$ basis.\} }\label{pbs}
\end{figure}

The 2-qubit codes in our protocol above can be robustly produced form SPDC process\cite{para}. 
In such a polarization space, $|0\rangle, |1\rangle$
are represented by  the horizontal polarization state and vertical polarization state.
In the ``classicalized'' protocol above, Bob need to take the operation of quantum parity check and quantum decoding
to the codes received from Alice. In practice, Bob can take the two operation together with a polarizing beam splitter
and obtain the result by post-selection. Bob's operation to the incoming 2-qubit code is shown in Fig.(\ref{pbs}). If he obtains
nothing in either D1 or D2 measurement, he aborts the code; if there is one photon on
each side of the PBS, he records the bit value and ``measurement basis''
according to the correspondence rule in step 2 in the protocol. Note that  a PBS transmits the horizontal qubits and reflects the vertical
qubits. In Alice's initial preparation of the quantum code, the two qubits have the same polarization. If one of them is flipped in
Z basis by the channel, the code will contain two different polarizations and both photons will be on the same
 side of the PBS, i.e., either D1 or D2 will be vacuum,  therefore the code will be discarded.
To verify the fact of one photon on each side of the PBS, Bob only needs to see that  both photon  detectors on each side of the
PBS click. Note that we only need  yes-no photon detectors here. The low efficiency of the detectors does not affect the security
of the protocol.
\section{6-state protocol}
Our protocol can obviously be extended to the 6-state protocol. In doing  so, Alice just change the initially random codes by
adding $N/4$ codes from $\{\frac{1}{2}[(|00\rangle+|11\rangle)\pm i(|00\rangle-\rangle|11\rangle)]\}$. This is equivalent to 
$\frac{1}{\sqrt 2}(\{|00\rangle
\mp i |11\rangle)\}$. She regards all this
type codes as Y bits. In decoding the codes, Bob's
``measurement basis'' is randomly chosen from 3 basis, with the basis of $\frac{1}{\sqrt 2}(|0\rangle\pm i|1\rangle)$ being added.
All decoded  X bits, Y bits and the same number of decoded Z bits which are randomly chosen will be used as the check bits. Since 
the Hadamard transformation
will switch the two eigenstates of $\sigma_y$, therefore in classicalized
protocol, after the decoding if Bob choose to take the measurement to B2 qubit
in Y basis, he shall records his ``measurement basis'' as Y but the 
``measurement outcome'' should be flipped in the Y basis so that to obtain 
everything the same as that in the 2-EPP with quantum storages. 
In such a way, if the channel
is symmetric, Bob will find $p_y\not= 0$. And he will know $p_x,p_y,p_z$
exactly instead of assuming $p_y=0$. This will increase the tolerable error rate accordingly. We find that the
tolerable bit flip rate and  phase flip rate are $30\%$ in such a case, i.e., our 6-state protocol can tolerate
a total flip rate of $45\%$ with the symmetric noisy channel.   
\section{Another viewpoint of our protocols}
To further clarify the concept of unconditional security with conditional advantage in efficiency of our protocols, here we view our 
scheme from a another viewpoint. We denote channel between Alice and Bob by operator $\hat C$, i.e. a transmitted qubit will be interacted 
with channel by  that operator.
We can now compare Chau protocol\cite{chau} with our protocol given the same 
channel $\hat C$.  In our protocol, if we can also regard the Alice's quantum encoding part and Bob's quantum
decoding part as part of the channel interaction. In such a case, our protocol with channel $\hat C$ is actually
just the Chau protocol\cite{chau} with a $lossy$ channel 
\begin{eqnarray}
\hat C' = \hat D \hat C \hat E
\end{eqnarray}    
where the operators $\hat E , \hat D$ represent for encoding and decoding respectively.
Note that so far all the known  secure protocols are also secure with a lossy channel, since
the purification scheme can be done with a lossy channel, with the device imperfections such as dark counting
being ignored here.  
The error rates for channel $\hat C'$ are functions of the error rates of channel $\hat C$. In particular,
they are related by Eq.(\ref{errorrate}) if errors caused by $\hat C$ are uncorrelated. To whatever channel,
the error rate should satisfy the same convergence condition to produce non zero bits of final key.
Specifically, we can use the following formula for the convergence condition:
\begin{eqnarray}
f(p_{x0},p_{y0},p_{z0})\le \eta \label{cc1}
\end{eqnarray} for channel $\hat C$
and 
 \begin{eqnarray}
f(P_{x},P_{y},P_{z})\le \eta\label{cc2}
\end{eqnarray} for channel $\hat C'$
where $(p_{x0},p_{y0},p_{z0})$ and $(P_{x},P_{y},P_{z})$ are detected error rates for channel $\hat C$
and channel $\hat C'$, respectively. $\eta$ is a small number, say $5\%$. Though the 
 above 2 equations have the same form, Eq.(\ref{cc2}) requires a different convergence condition to $(p_{x0},p_{y0},p_{z0})$. If the error rates
of Channel $\hat C'$ and that of channel $\hat C$ are related by eq.(\ref{errorrate}), then we can solve
the condition of formula (\ref{cc2}) numerically and obtain the thresholds to  $(p_{x0},p_{y0},p_{z0})$, which is 
just the claimed threshold in our protocols.

Roughly speaking, our protocol is equivalent to the one which first investigates the channel properties and then
find a scheme to reduce the s-channel noise based on the investigated prpoerties and then carry out Chau protocol\cite{chau}.
The channel noise is not unconditionally reduced, it is based on the investigated properties of the 
s-channel itself. At the time we are really carrying out  QKD, Eve. may attack it and the assumed properties is not 
necessarily true. But we still believe that  the protocol will work in the way as it is assumed in most of the cases
for the reasons we have addressed before. It's a reasonable assumption 
that it should be an advantage if we have a way
to decrease the noise of the s-channel, even though in principle Eve. may do 
everything to the channel in QKD.
\section{ Concluding remark.}
We have proposed an entanglement assisted protocol for quantum key distribution. The protocol is unconditionally secure with a conditional advantage of 
an improved threshold of the channel error rate. 
Given the uncorrelated noise of s-channel, in 4-state case, the tolerable 
bit flip or phase flip rate is $26\%$; 
in the  6 state case, the values are increased to $30\%$, if the noise is symmetric.
The initial 2-qubit entangled state can be
produced through SPDC process\cite{para}. Bob's quantum parity check and decoding can be done with a 
polarizing beam splitter. Obviously, all devices  appeared in our proposed scheme are 
just ordinary devices in linear optics.

There are also some loosen ends for our protocol in practical realization. In the SPDC process, there are also some multi-pair emission with small probability.
This type of emission will affect the security that should be taken into consideration for the security proof for a practical
entanglement assisted QKD protocol with SPDC process. We believe this can be solved in the similar idea for the case
of imperfection of the single photon source for the BB84 scheme. We have neither considered the devices imperfections, e.g.,
the dark counting of the photon detectors. Neither have we considered the Trojan Horse attack to the protocol. 
Similar to other QKD proposals\cite{gl,chau}, here the tolerable upper bound for channel 
error rate is calculated
asymptotically. In practice, the number of qubits sent to Bob is always finite.
With the statistical fluctuation being taken into consideration, the tolerable
flip rate of the channel is decreased in practice with a finite number of initial qubits.

\newpage
{\bf Acknowledgement:} I thank Prof Imai H for support. 


\begin{thebibliography}{99}
\bibitem{BB}
C. H. Bennett and G. Brassard, 
``Quantum cryptography: Public-key distribution and coin tossing,'' 
in {\em Proceedings of IEEE International Conference on Computers, 
Systems and Signal Processing, Bangalore, India, 1984},  (IEEE Press,
1984), pp. 175--179;
C.H. Bennett and G. Brassard,
``Quantum public key distribution,''
IBM Technical Disclosure Bulletin {\bf 28}, 3153--3163 (1985).
\bibitem{gisin} N. Gisin, G. Ribordy, W. Tittel, and H. Zbinden,
``Quantum Cryptography,'', Reviews of Modern Physics, vol. 74, pp. 145-195.
Also [Online]
Available: http://xxx.lanl.gov/abs/quant-ph/0101098.
\bibitem{qkd} H.-K.~Lo and H.~F.~Chau, ``Unconditional security of
quantum key distribution over arbitrarily long distances,'' Science,
283, 2050(1999)
\bibitem{6state}D. Bruss, Phys. Rev. Lett. 81, 3018(1998).
\bibitem{para} P. G. Kwiat, K. Mattle, H. Weinfurter, A. Zeilinger, A.V. Sergienko, and Y. H. Shih, Phys. Rev. Lett. 75, 4337(1995).
\bibitem{gl} D. Gottesman and H.-K. Lo, IEEE Transactions on
 Information Theory, 49, 457(2003), quant-ph/0105121, ``Proof of
security of quantum key distribution with two-way classical communication''.
\bibitem{mayersqkd} D. Mayers, ``Unconditional security in
Quantum Cryptography,'' Journal of ACM, vol. 48, Issue 3, p. 351-406. 
Also [Online] Available:
http://xxx.lanl.gov/abs/quant-ph/9802025.
\bibitem{others}
E. Biham, M. Boyer, P. O. Boykin, T. Mor, and V. Roychowdhury,
``A proof of the security of quantum key distribution,''
in {\it Proceedings of
the Thirty-Second Annual ACM Symposium on Theory of Computing} (STOC)
(ACM Press, New York, 2000), p. 715.
\bibitem{others2}
M. Ben-Or, Unpublished.
\bibitem{BDSW} C. H. Bennett, D. P. DiVincenzo, J. A. Smolin,
and W. K. Wootters, ``Mixed state entanglement and quantum error
correction,'' {\it Phys. Rev.}, vol. A54, 3824, 1996.
\bibitem{bdswa}
C. H. Bennett, D. P. DiVincenzo, J. A. Smolin and W. K. Wootters,
``Mixed state entanglement and quantum error correction,''
Phys. Rev. A,  {\bf 54}, 3824--3851 (1996),
{\em \mbox{arXive} e-print} quant-ph/9604024.
\bibitem{deutsch} D.~Deutsch, A.~Ekert, R.~Jozsa, C.~Macchiavello,
S.~Popescu, and A.~Sanpera, ``Quantum privacy amplification and the
security of quantum cryptography over noisy channels,''
{\it Phys.~Rev.~Lett.}, vol.  77, p. 2818, 1996.
Also, [Online] Available:
http://xxx.lanl.gov/abs/quant-ph/9604039.  Erratum
Phys.~Rev.~Lett. {\bf 80}, 2022 (1998).
\bibitem{shorpre} P. W. Shor and J. Preskill, ``Simple proof of
security of the BB84 quantum key distribution protocol,''
{\it Phys. Rev. Lett.}, vol. 85, p. 441, 2000. Also,
[Online] Available: http://xxx.lanl.gov/abs/quant-ph/0003004.
\bibitem{squeezed} D. Gottesman and J. Preskill, ``Secure
quantum key distribution using squeezed states,''
{\it Phys. Rev.}, vol. A63, p. 22309, 2001.
\bibitem{CSS}
A. R. Calderbank and P. Shor, 
``Good quantum error correcting codes exist,''
Phys. Rev. A {\bf 54}, 1098--1105 (1996),
{\em \mbox{arXive} e-print} quant-ph/9512032; 
A. M. Steane, 
``Multiple particle interference and error correction,'' 
Proc. R. Soc. London A
{\bf 452}, 2551--2577 (1996),
{\em \mbox{arXive} e-print} quant-ph/9601029.
\bibitem{chau} H. F. Chau, in quant-ph/0206050, ``Practical scheme to share a secret key
through up to 27.6
\bibitem{pad} H. K. Lo, quant-ph/0201030, ``Method for decoupling
error correction from privacy amplification''.
\bibitem{sremark} We do not have to require that the detected error rate
satisfy or approximately satisfy Eq.(\ref{errorrate}) as the condition
of an acceptable error rate. Even the detected error rates are significantly
larger than the expected values given by Eq.(\ref{errorrate}), they can still
continue the protocol. In such a case they will obtain a much shorter final
key to which Eve's information is exponentially close to zero.
In any case they can carry out the whole protocol and accept the final key
provided that they believe the final key in that length deserves the cost
of carrying out the protocol. 
\bibitem{notek} The ``preparation basis'' here 
is defined by the basis of the original qubit which corresponds 
the 2-qubit code by our encoding method.
\bibitem{note0} Here Bob has  omitted  the Hadamard transformation as 
appeared in the EPP, therefore 
the measurement basis to B2 and Bob's recorded `` measurement basis'' 
to the decoded qubit are different.
\end{thebibliography}
\end{document}